\documentclass[aps,reprint,longbibliography,superscriptaddress]{revtex4-2}
\usepackage{mathrsfs}
\usepackage{amsmath,gensymb}
\usepackage{amsfonts}
\usepackage{amssymb}
\usepackage{amsthm}
\usepackage{graphicx}
\usepackage{natbib}
\usepackage{xcolor}
\usepackage{hyperref}
\usepackage{bm}
\usepackage[caption=false]{subfig}
\usepackage{verbatim}
\usepackage[normalem]{ulem}

\usepackage{xtab,afterpage}
\usepackage{lipsum}

\begin{document}

\title{Inverse proximity effect in thin-film superconductor/magnet heterostructures with metallic and insulating magnets}

\author{V. A. Bobkov}
\affiliation{Moscow Institute of Physics and Technology, Dolgoprudny, 141700 Moscow region, Russia}

\author{G. A. Bobkov}
\affiliation{Moscow Institute of Physics and Technology, Dolgoprudny, 141700 Moscow region, Russia}

\author{I. V. Bobkova}
\affiliation{Moscow Institute of Physics and Technology, Dolgoprudny, 141700 Moscow region, Russia}
\affiliation{National Research University Higher School of Economics, 101000 Moscow, Russia}

\begin{abstract}
Proximity effect in thin-film superconductor (S)/magnet heterostructures with different types of magnets including ferromagnets, antiferromagnets and altermagnets is widely considered in the framework of an effective model, where the heterostructure is replaced by a homogeneous superconductor in the presence of a homogeneous exchange field $h_{\rm eff}$ of a corresponding type. Here we study the extent to which such a model is actually applicable to ballistic thin-film superconductor/magnetic heterostructures. In particular, a comparative analysis of thin-film superconductor/magnetic metal and superconductor/magnetic insulator heterostructures is performed. Metallic and insulating ferromagnets (FM, FI) and altermagnets (AM, AI) are considered. It is shown that in the S/FI and S/AI heterostructures the the proximity effect creates a well-defined spin splitting
of the electronic spectra in the S layer. Thus, they are well described by the effective model. At the same time, the
proximity effect in S/FM and S/AM heterostructures also creates a
spin splitting of the spectra of the S layer, but it has a chaotic spectral and spatial distribution and unpredictable amplitude and, in general, cannot be detected via the spin splitting of the superconducting density of states. Thus,
the effective model is not applicable to such heterostructures.  
Nevertheless, we demonstrate that they support well-pronounced triplet
correlations and, thus, can be used for spintronics applications. 
\end{abstract}

\maketitle

\section{Introduction}
\label{intr}

One of the key effects underlying the applications of superconducting spintronics is the proximity effect in superconductor/ferromagnet (S/F) heterostructures. The direct proximity effect is the penetration of superconducting correlations into a ferromagnet (if it is metallic) with partial conversion of singlet correlations into triplet ones \cite{Buzdin2005,Bergeret2005}. There is also an inverse proximity effect, which is a modification of the electronic properties of a superconductor as a result of contact with a ferromagnet. This modification occurs on scales of the order of the superconducting coherence length $\xi_S$ from the S/F interface and therefore most significantly affects the properties of thin-film superconductors with a thickness less than $\xi_S$. The inverse proximity effect has two most important manifestations. The first one is a partial conversion of singlet superconducting correlations into triplet ones, and the second is the spin splitting of the local electronic density of states (LDOS) in the superconductor \cite{Bergeret2018,Heikkila2019}. In thin-film superconductors the spin-splitting is very similar to the Zeeman splitting, which occurs under the applied in-plane magnetic field. 

Both manifestations of the inverse proximity effect play a very important role in superconducting spintronics \cite{Linder_review,Eschrig_review,Bergeret2018}. For example, the emergence of triplet correlations as a result of singlet-triplet conversion is the basis of the superconducting spin-valve effect \cite{DEGENNES1966,Li2013,Oh1997,Jara2014,Singh2015,Kamashev2019,Tagirov1999,Fominov2003,Zhu2010,Moraru2006,Singh2007,Banerjee2014,Leksin2012,Westerholt2005,Deutscher1969,Gu2002,Gu2015,Karminskaya2011,Wu2014,Halterman2015,Moen2017,Moen2018,Moen2020,Valls_book}, and the spin splitting of the LDOS underlies the effects of superconducting spin caloritronics \cite{Bergeret2018,Heikkila2019,Bobkova2017,Bobkova2021} and, in particular, the giant thermoelectric effect \cite{Machon2013,Ozaeta2014,Machon2014,Kolenda2016,Kolenda2017}. Until now, it was considered in the literature that thin-film superconductors in contact with a ferromagnet can be described within the framework of an effective model. In this model the heterostructure is replaced by a homogeneous superconductor in some effective Zeeman field $h_{\rm{eff}}$ \cite{Bergeret2001,Cottet2009,Eschrig2015} and, possibly, with the additional influence of some effective depairing parameter responsible for the leakage of correlations into the ferromagnet (if it is metallic) or acting as magnetic impurities in the case of an interface with a ferromagnetic insulator \cite{Eschrig2015}. In the framework of the effective model $h_{\rm{eff}}$ is a generator of both triplet correlations and spin splitting of the LDOS in the superconductor. Following this logic, there is a direct connection between the presence or absence of spin splitting of the LDOS, which is accessible for direct experimental measurement, and the presence or absence of triplet correlations in the superconductor.

Spin splitting of the LDOS has been demonstrated several times in thin-film superconductor/ferromagnetic insulator (S/FI) heterostructures \cite{Hao1990,Moodera1988,Tedrow1986,Meservey1994,Kolenda2017,Strambini2017,Bergeret2018,Heikkila2019}. However, spin splitting of the LDOS has not yet been experimentally implemented in thin-film superconductor/ferromagnetic metal (S/FM) heterostructures. Only signatures of the minigap spin splitting in S/F/N/S Josephson  junctions were reported \cite{Golikova2012}. The question arises whether this means that the proximity with a ferromagnetic metal does not induce an effective exchange field and triplet correlations in the superconductor. Moreover, a more general question can be raised: does the absence of experimentally observed splitting of peaks in the LDOS always mean the absence of spin splitting of the electron spectra of the superconducting film and the absence of triplet correlations? The present work is devoted to the study of this issue.

A comparative analysis of thin-film S/FM and S/FI heterostructures is performed. We show that in the case of S/FM heterostructures the proximity effect is usually strong. The spin splitting of the electronic spectra in the superconducting film is also typically large. However, unlike S/FI heterostructures, in general this proximity effect does not lead to a clear splitting of the LDOS peaks. That is, the experimentally observed splitting of the LDOS cannot serve as a criterion of existence and a quantitative measure of triplet correlations in a superconductor. Thus, the widely used model that describes a S/F thin-film bilayer as a homogeneous superconductor in an effective Zeeman field $h_{\rm{eff}}$ is not adequate. Further, it is clearly demonstrated that S/FM heterostructures, in which no splitting of the LDOS is observed, can be promising from the point of view of superconducting spintronics and exhibit a significant spin-valve effect. In addition to S/F heterostructures, we also investigate the proximity effect in superconductor/altermagnet heterostructures, which are currently being actively studied theoretically \cite{Sun2023,Zhang2024,Banerjee2024,Li2023,Papaj2023,Ouassou2023,Beenakker2023,Chourasia2025,Maiani2025,Vasiakin2025}. The applicability of the analogous effective model in which the real heterostructure is replaced by a superconductor in an effective altermagnetic exchange field is investigated.

The paper is organized as follows. In Sec.~\ref{sec:model} we describe systems under study and model assumptions we use for theoretical description of the systems. In Sec.~\ref{sec:theory} we provide a theoretical analysis of the proximity-induced Zeeman splitting in thin metallic films. Sec.~\ref{sec:numerical} is devoted to a comparative numerical demonstration of the triplet correlations, Zeeman splitting of the electronic spectra and LDOS for S/FM and S/FI heterostructures and spin valve effect in F/S/F structures. In Sec.~\ref{sec:numerical_AM} we present numerical results for S/AM heterostructures. Our conclusions are summarized in Sec.~\ref{sec:conclusions}.

\section{System and model}
\label{sec:model}

\begin{figure}[tbh!]
\centering
\includegraphics[width=0.85\columnwidth]{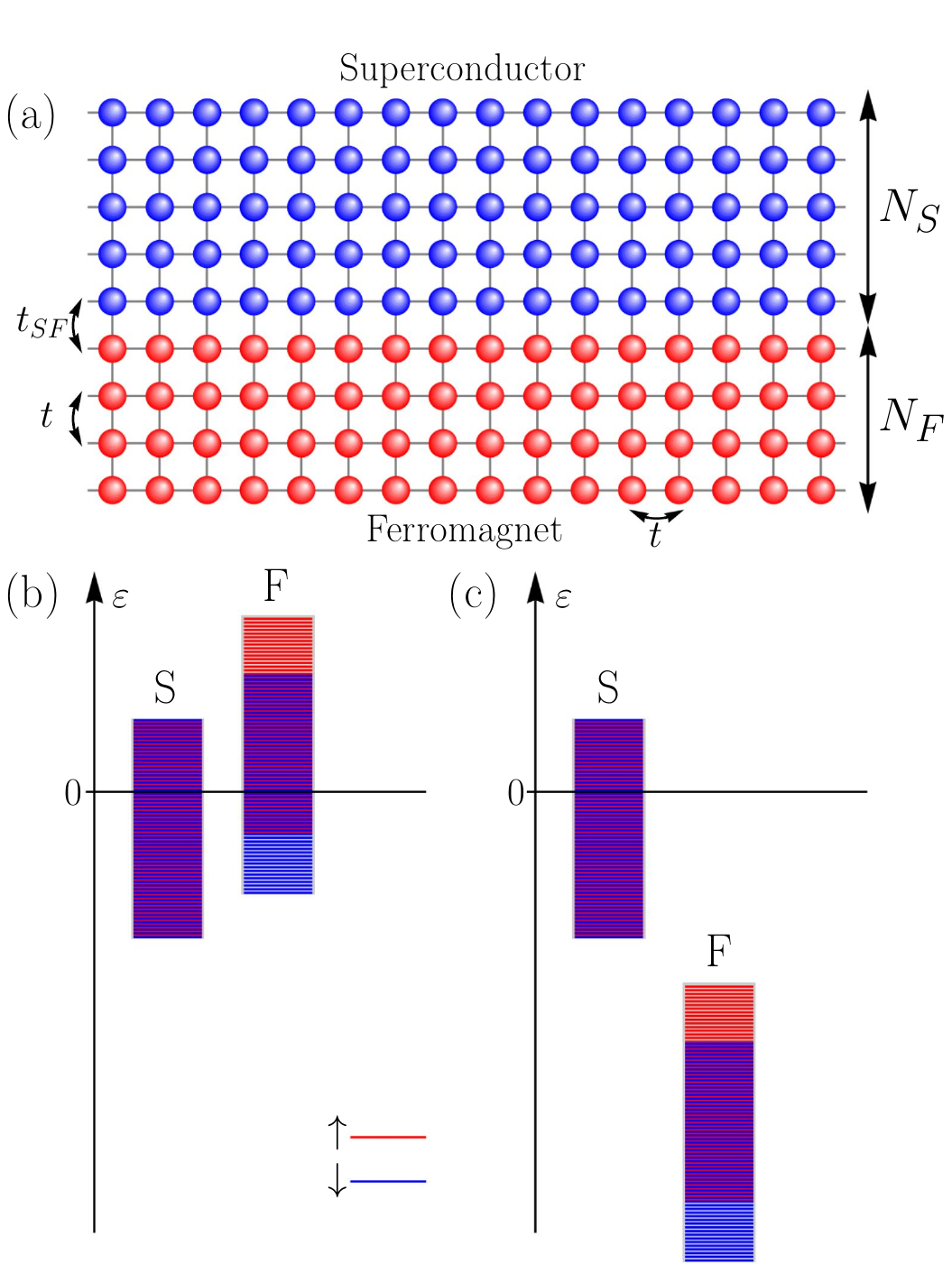}
\caption{(a) Sketch of the S/F bilayer under consideration. Intralayer nearest-neighbor hopping parameter $t$ (the same for the S and F layers) and interlayer hopping parameter $t_{SF}$ are shown. The system is infinite in the $(x,y)$-plane. (b)-(c) Sketches of the superconductor and ferromagnet conduction bands for the case of S/FM bilayer (b) and S/FI bilayer (c). Red and blue colors correspond to spin-up and spin-down density of states, respectively. The bottom of the conduction band in the S(F) layer is determined by the parameter $\mu_{S(F)}$.} 
 \label{fig:sketch}
\end{figure}

An example of a thin-film heterostructure under study is sketched in Fig.~\ref{fig:sketch}(a). All studied heterostructures consist of $N_F$ infinite in the $(x,y)$-plane layers of a magnetic material, which can be FI, FM or AM, interfaced with $N_S$ layers of a superconducting material. The S/F systems are described by the following tight-binding Hamiltonian on a cubic lattice:
\begin{align}
\hat H= &- \sum \limits_{ \langle \bm i \bm j \rangle \alpha} t_{\bm i \bm j}\hat \psi_{\bm i \alpha}^{\dagger} \hat \psi_{\bm j \alpha} + \sum \limits_{\bm i } (\Delta_{\bm i} \hat \psi_{\bm i\uparrow}^{\dagger} \hat \psi_{\bm i\downarrow}^{\dagger} + H.c.) \nonumber \\
&- \sum \limits_{ \bm i \alpha} \mu_{\bm i} \hat \psi_{\bm i \alpha}^{\dagger} \hat \psi_{\bm i \alpha} + \sum \limits_{\bm i,\alpha \beta} \hat \psi_{\bm i\alpha}^{ \dagger} (\bm h_{\bm i} \bm \sigma)_{\alpha \beta} \hat \psi_{\bm i\beta},
\label{eq_hamBDG}
\end{align}
where $\hat{\psi}_{\bm i \sigma}^{\dagger}(\hat{\psi}_{\bm i \sigma})$ is the creation (annihilation) operator for an electron with spin $\sigma=\uparrow,\downarrow$ at the site $\bm i$. We only consider nearest-neighbor hopping with the hopping amplitude $t_{\bm i \bm j}=t$ for the intralayer hopping (the same in the superconducting and magnetic layers) and $t_{\bm i \bm j}=t_{SF}$ for the interlayer hopping at the superconductor/magnet interface. $\langle \bm i \bm j \rangle $ means summation over the nearest neighbors. $\mu_{\bm i}$ is the on-site energy at the site $\bm i$, which equals to $\mu_{S(F)}$ in the superconducting (magnetic) part of the structure. $\Delta_{\bm i}$ accounts for on-site s-wave pairing. It is only nonzero in the S layer. $h_{\bm i}$ is the exchange field acting on the conductivity electrons at the site $\bm i$ belonging to the ferromagnetic layer. $ \bm \sigma = (\sigma_x,\sigma_y,\sigma_z)^T$ is the vector of Pauli matrices in spin space. The S/FI and S/FM heterostructures differ by the value of $\mu_F$, which controls the energy of the bottom of the conduction band in the ferromagnet, see Fig.~\ref{fig:sketch}(b). 

S/AM heterostructures are described by the Hamiltonian (\ref{eq_hamBDG}) with a modified magnetic term:
\begin{align}
\sum \limits_{\bm i,\alpha \beta} \hat \psi_{\bm i\alpha}^{ \dagger} (\bm h_{\bm i} \bm \sigma)_{\alpha \beta} \hat \psi_{\bm i\beta} \to \sum \limits_{ \langle \bm i \bm j \rangle \alpha} \hat \psi_{\bm i \alpha}^{\dagger} (\bm h_{\bm i \bm j} \bm \sigma)_{\alpha \beta} \hat \psi_{\bm j\beta},
\label{eq:ham_am}
\end{align}
where we set $\bm h_{\bm i \bm j} = h \bm e_z$ if $\bm j = \bm i \pm \bm e_x$ and $\bm h_{\bm i \bm j} = -h \bm e_z$ if $\bm j = \bm i \pm \bm e_y$. In homogeneous materials this corresponds to a low-energy exchange field $h (k_x^2-k_y^2)\sigma_z$.

We assume that our structures are in the ballistic limit. For thin-film heterostructures with thicknesses of each layer of the order of several nm this approximation is reasonable for electron motion in the direction normal to the interface. At the same time, the impurity scattering can be important for motion in plane of the structure. We discuss the influence of impurity scattering on the obtained results in the following sections.

\section{Analysis of the proximity-induced Zeeman splitting of the electronic spectra in thin metallic films}
\label{sec:theory}

\begin{figure}[tbh!]
\centering
\includegraphics[width=0.85\columnwidth]{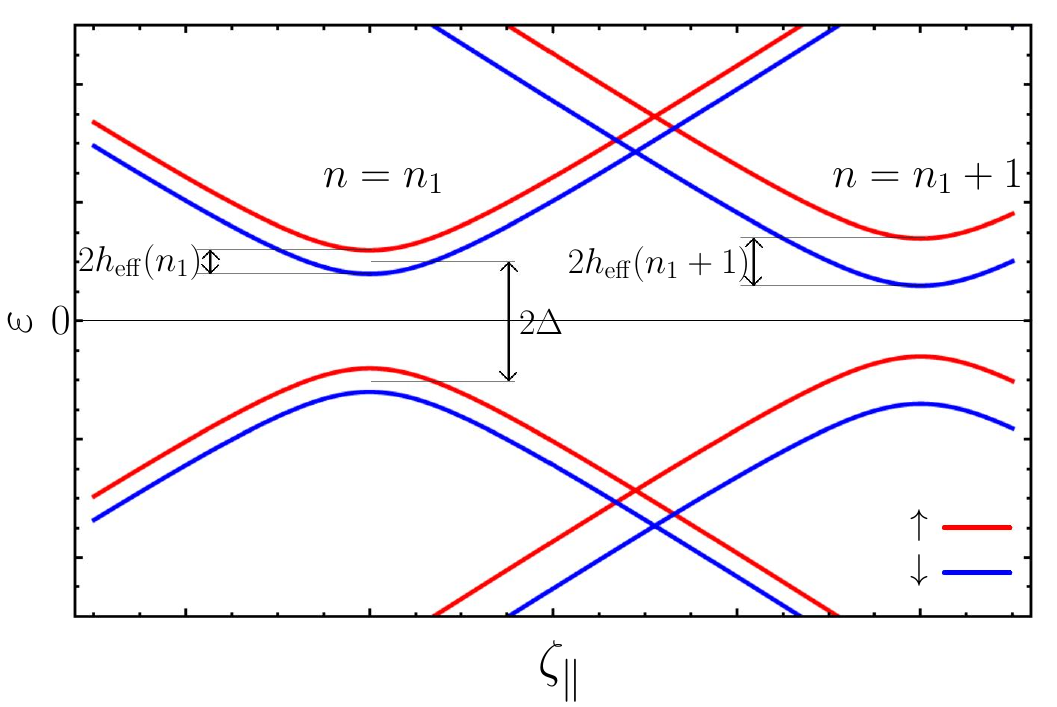}
\caption{Two discrete spin-split branches of the S layer spectrum $\varepsilon_S^{\uparrow,\downarrow} (\zeta_\parallel, n)$. Spin-splitting of each of the branches, which is equal to $2 h_{\rm{eff}}(n)$ is shown.} 
 \label{fig:h_eff_sketch}
\end{figure}

In this section it is enough to consider the S layer in its normal state corresponding to $\Delta = 0$. Let the $z$ axis be directed perpendicular to the interface and the $z$-coordinates of the superconducting layers be equal to $-(N_S-1)a$, $-(N_S-2)a$,..., $0$ and $z$-coordinates of the ferromagnetic layers be equal to $a$, $2a$,..., $N_Fa$, where $a$ is the lattice constant.  We can write the wave function in the S and F layers as the sum of the incident and reflected waves:
\begin{align}
    {\rm S:}~~~~\psi_\sigma(\bm r)=A_S^\sigma e^{i(k_x x+k_y y)}(e^{ik_S^\sigma z}+r_S^\sigma e^{-ik_S^\sigma z}), \label{wave_S} \\
    {\rm F:}~~~\psi_\sigma(\bm r)=A_F^\sigma e^{i(k_x x+k_y y)}(e^{ik_F^\sigma z}+r_F^\sigma e^{-ik_F^\sigma z}),
    \label{wave_F}
\end{align}
where $\bm r=(x,y,z)^T$ is the discrete radius-vector of a given site and for brevity we have designated $k_z$ in the S(F) layer as $k_{S(F)}$. The Schr\"odinger equation for $\psi_\sigma(\bm r)$ at $\Delta=0$ takes the form: 
\begin{align}
    \label{eq:Schrodinger}
    -t\sum_{s=\pm 1}[ \psi_\sigma (x+sa,y,z)+\psi_\sigma (x,y+sa,z)+ \\ \nonumber
    +\psi_\sigma (x,y,z+sa)]-\tilde\mu_{S(F)}^\sigma(\bm r)\psi_\sigma(\bm r)=\varepsilon^\sigma \psi_\sigma(\bm r),
\end{align}
where $\tilde\mu_{S(F)}^\sigma(\bm r)=\mu_{S(F)}(\bm r)- h(\bm r)\sigma$. Taking Eq.~(\ref{eq:Schrodinger}) at a site $\bm r$, which nearest neighbors belong to the same material (S or F layer) and substituting Eqs.~(\ref{wave_S}) and (\ref{wave_F}) into Eq.~(\ref{eq:Schrodinger}), we obtain the dispersion relation for the S and F layers:
\begin{align}
    \zeta_{\parallel}-2t\cos(k_{S(F)}^\sigma a)-\tilde\mu^\sigma_{S(F)}=\varepsilon^\sigma,
    \label{eq:homogeneous}
\end{align}
where $\zeta_{\parallel}=-2t[\cos(k_x a)+\cos(k_y a)]$. At the edges of the sample $z=N_Fa$ and $z=-(N_S-1)a$ Eq.~(\ref{eq:Schrodinger}) is modified: there is no hopping term to the site $z\pm a$ for right and left edge, respectively. The equations at sites $z=0,a$ are also modified due to the hopping at sites belonging to the different layer. Substituting wave functions in the form of Eqs.~(\ref{wave_S}) and (\ref{wave_F}) into Eq.~(\ref{eq:Schrodinger}) taken at $z=-(N_S-1)a,0,a,N_Fa$ we obtain the following relations: 
\begin{align}
     &A_S^\sigma(\zeta_{\parallel}-\mu_S-\varepsilon^\sigma) (e^{ik_S^\sigma a (1-N_S)}+r_S^\sigma e^{ik_S^\sigma a (N_S-1)})- \nonumber \\ 
     &t A_S^\sigma(e^{ik_S^\sigma a (2-N_S)}+r_S^\sigma e^{ik_S^\sigma a(N_S-2)})=0, \label{bc01} \\ 
     &A_S^\sigma[(\zeta_{\parallel}-\mu_S-\varepsilon^\sigma) (1+r_S^\sigma)-t (e^{-ik_S^\sigma a }+r_S^\sigma e^{ik_S^\sigma a})]- \nonumber \\ 
     &t_{SF} A_F^\sigma(e^{ik_F^\sigma a }+r_F^\sigma e^{-ik_F^\sigma a})=0, \label{bc02} \\ 
     &A_F^\sigma[(\zeta_{\parallel}-\tilde\mu^\sigma_F-\varepsilon^\sigma) (e^{ik_F^\sigma a}+r_F^\sigma e^{-ik_F^\sigma a})- \nonumber \\ 
     &t (e^{2ik_F^\sigma a }+r_F^\sigma e^{-2ik_F^\sigma a})]-t_{SF} A_S^\sigma(1+r_S^\sigma)=0, \label{bc03} \\ 
     &A_F^\sigma(\zeta_{\parallel}-\tilde\mu^\sigma_F-\varepsilon^\sigma) (e^{ik_F^\sigma a N_F}+r_F^\sigma e^{-ik_F^\sigma a N_F})- \nonumber \\ 
     &t A_F^\sigma(e^{ik_F^\sigma a (N_F-1)}+r_F^\sigma e^{ik_F^\sigma a(1-N_F)})=0 .
     \label{bc04}
\end{align}
Subtracting from Eqs.~(\ref{bc01})-(\ref{bc02}) the homogeneous relation Eq.~(\ref{eq:homogeneous}) taken in the S layer, and from Eqs.~(\ref{bc03})-(\ref{bc04}) the homogeneous relation Eq.~(\ref{eq:homogeneous}) taken in the F layer, we obtain the following boundary conditions:
\begin{align}
    &(e^{-ik_S^\sigma d_S}+r_S^\sigma e^{ik_S^\sigma d_S})A_S^\sigma=0, \nonumber \\ 
    &(e^{ik_F^\sigma (d_F+a)}+r_F^\sigma e^{-ik_F^\sigma (d_F+a)})A_F^\sigma=0, \nonumber \\ 
    &(e^{ik_S^\sigma a}+r_S^\sigma e^{-i k_S^\sigma a})A_S^\sigma=\frac{t_{SF}}{t} (e^{ik_F^\sigma a}+r_F^\sigma e^{-ik_F^\sigma a})A_F^\sigma, \nonumber \\ 
    &(1+r_F^\sigma)A_F^\sigma=\frac{t_{SF}}{t}(1+r_S^\sigma)A_S^\sigma .
   \label{bc}
\end{align}
where $d_{S(F)}=N_{S(F)}a$. From Eqs.~(\ref{bc}), which represent the linear system of equations for the unknown coefficients $A_{S,F}^\sigma$, $r_{S,F}^\sigma$, we obtain the following equation for the relationship between $k_S^\sigma$ and $k_F^\sigma$:
\begin{align}
    \frac{\sin (k_S^\sigma (d_S+a))\sin (k_F^\sigma (d_F+a))}{\sin (k_S^\sigma d_S)\sin (k_F^\sigma d_F)}=\frac{t_{SF}^2}{t^2} .
    \label{kS_kF1}
\end{align}
The second required equation for determination of $k_{S}^\sigma$ and $k_{F}^\sigma$ one can obtain by substracting Eqs.~(\ref{eq:homogeneous}) for the S and F layers:
\begin{align}
    -2t(\cos(k_S^\sigma a)-\cos(k_F^\sigma a))=\mu_S-\tilde\mu^\sigma_F .
 \label{kS_kF2}   
\end{align}
Eqs.~(\ref{kS_kF1}), (\ref{kS_kF2}) allow us to calculate discrete values of the momentum component $k_{S,F}^\sigma$ perpendicular to the interface. Then the electron spectrum in the S layer $\varepsilon_{S}^\sigma(\zeta_\parallel, n)$ can be obtained from Eq.~(\ref{eq:homogeneous}). Here $n$ is the number of the discrete perpendicular momentum component $k_S^\sigma(n)$.  Our goal is to find the spin splitting $\varepsilon_S^\uparrow (\zeta_\parallel, n) - \varepsilon_S^\downarrow (\zeta_\parallel, n)$ of the S layer spectra induced by the proximity to the F layer, which is the value of the doubled effective exchange field $h_{\rm{eff}}(n)$ induced by the ferromagnet in the superconductor. The example of two spin-split discrete branches of the S layer spectrum is presented in Fig.~\ref{fig:h_eff_sketch}.

Let us consider two limiting cases when an analytical solution for $h_{\rm{eff}}$ can be found explicitly. First, we write down the trivial solutions for individual layers at $t_{SF}=0$:
\begin{align}
    k_{S(F)}^0(n)=\frac{\pi n}{d_{S(F)}+a},~~~n\in [1, N_{S(F)}], \nonumber \\ 
    \varepsilon_{S(F)}^{\sigma,0}(n)=-\tilde \mu^\sigma_{S(F)}+\zeta_\parallel - 2t\cos[k_{S(F)}^0(n) a].
    \label{spectrum_ind}
\end{align}
The first considered limiting case is an insulating ferromagnet, i.e. when $\tilde \mu_F^\sigma-\mu_S\gg t$. In this case $k_F^\sigma$ is imaginary and for a given $n$ Eq.~(\ref{kS_kF1}) can be written as:
\begin{align}
    \frac{\sin (k_S^\sigma (d_S+a))}{\sin (k_S^\sigma d_S)}e^{\kappa_F^\sigma a}=\frac{t_{SF}^2}{t^2}.
    \label{kS_kF1_lim1}
\end{align}
where $\kappa_F^\sigma=ik_F^\sigma$ and from Eq.~(\ref{kS_kF2}) one can obtain
\begin{align}
    e^{\kappa_F^\sigma a}=2(\cos (k_S^0a)+\frac{\mu_S-\tilde\mu^\sigma_F}{2t}).
    \label{kS_kF2_lim1}
\end{align}
When writing Eqs.~(\ref{kS_kF1_lim1})-(\ref{kS_kF2_lim1}), we took into account that $e^{\kappa_F^\sigma a} \gg 1$. Substituting Eq.~(\ref{kS_kF2_lim1}) into Eq.~(\ref{kS_kF1_lim1}) and expanding by the first order with respect to the small parameter $t/(\tilde \mu_F^\sigma-\mu_S)$ we obtain:
\begin{align}
    \Delta k^\sigma=k_S^\sigma-k^0_S=\frac{t_{SF}^2 \sin(k_S^0 d_S)(-1)^n}{(d_S+a)t^2(2\cos(k_S^0a)+(\mu_S-\tilde\mu^\sigma_F)/t)}.
    \label{kS_lim1}
\end{align}
Then we can find the effective exchange field $h_{\rm{eff}}$ for a given branch $n$:
\begin{align}
    h_{\rm eff}(n)=\frac{1}{2}[\varepsilon^\uparrow(\zeta_\parallel,n)-\varepsilon^\downarrow (\zeta_\parallel,n)] .
    \label{h_eff_def}
\end{align}
Assuming that $h\ll \mu_F-\mu_S$, from Eqs.~(\ref{h_eff_def}), (\ref{kS_lim1}) and (\ref{eq:homogeneous}) one can obtain:
\begin{align}
    &h_{\rm eff}(n)=at\sin(k_S^0 a)(\Delta k^\uparrow-\Delta k^\downarrow)=\nonumber \\  &-\frac{2 a \sin(k_S^0 a)\sin (k_S^0 d_S)t_{SF}^2 h(-1)^n}{(d_S+a)(\mu_S-\mu_F)^{2}}=\nonumber  \\  &\frac{2 a t_{SF}^2h\sin^2(\frac{\pi n}{N_S+1})}{(d_S+a)(\mu_F-\mu_S)^{2}}.
    \label{h_eff_lim1}
\end{align}
In the framework of the considered model the spin splitting of each branch does not depend on energy. However, in more general case $h_{\rm{eff}}$ can also depend on energy. Then if we are interested in superconducting phenomena we need to take the energy near the Fermi surface ($\varepsilon\sim 0$).

Another way to describe the influence of the ferromagnetic insulator on the  superconductor is via the quasiclassical theory with boundary conditions formulated in terms of a spin-dependent interfaces scattering matrix \cite{Tokuyasu1988,Millis1988}, which results in the same physical conclusions.
In such framework spin-mixing angle $\phi_h$, which is the phase difference for electrons with spins up and down reflected from a ferromagnet, is defined as:
\begin{align}
    \phi_h={\rm Arg}(r_S^{*\uparrow})-{\rm Arg}(r_S^{*\downarrow}),
\end{align}
where $r_S^{*\sigma}=r_S^\sigma e^{-2i k_S^\sigma a} $. Then for the considered model it takes the form:
\begin{align}
    \phi_h=2(d_S+a)(\Delta k^\uparrow- \Delta k^\downarrow).
\end{align}

\begin{figure}[tbh!]
\centering
\includegraphics[width=0.9\columnwidth]{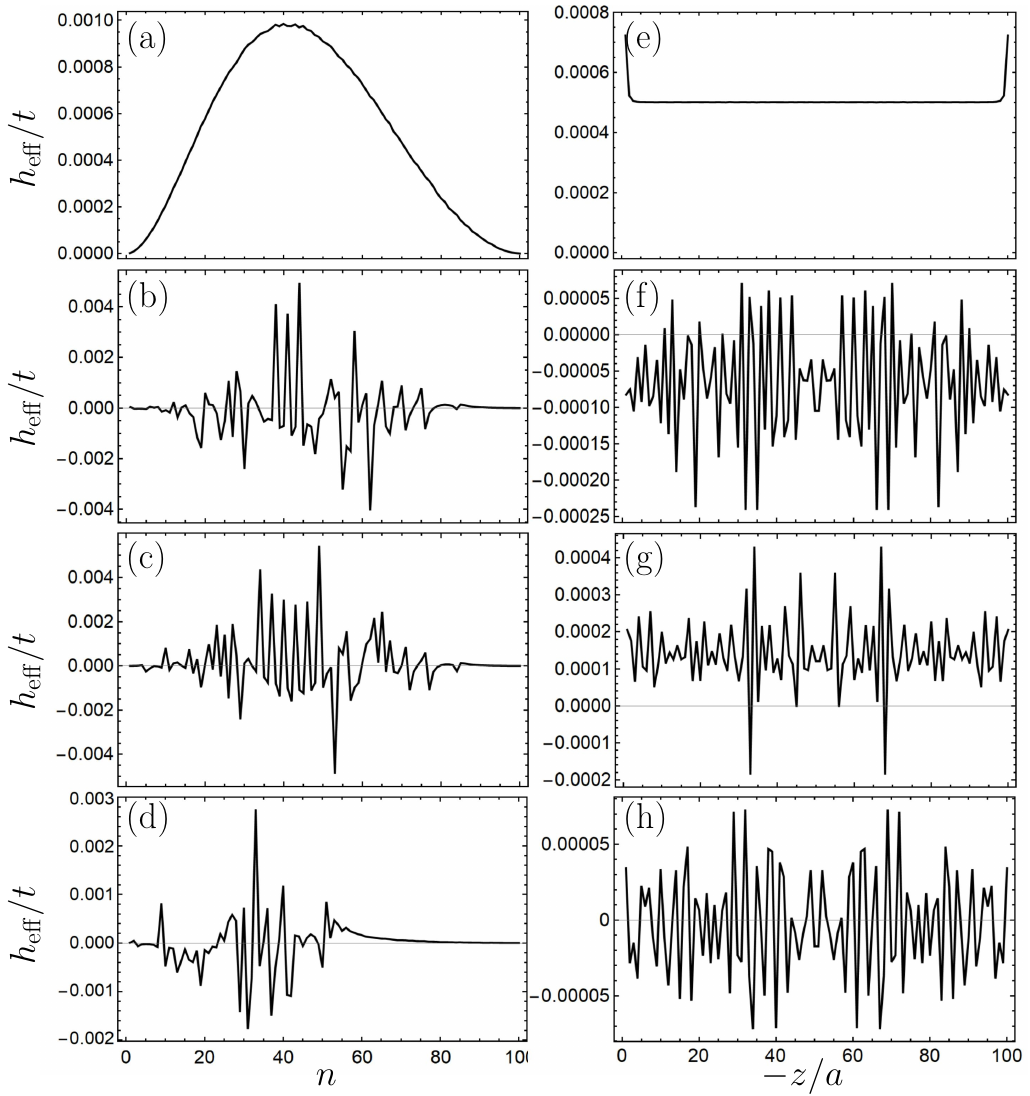}
\caption{Left column: dependence of $h_{\rm{eff}}$ on the branch number $n$ for different S/F heterostructures.  Right column: spatial dependence of $h_{\rm{eff}}$ on the number of the superconducting atomic layer for a corresponding heterostructure. (a),(e) S/FI heterostructure with $\mu_F=-4t$, $h=t$, $t_{SF}=t$, $N_F=70$; (b),(f) S/FM heterostructure with $\mu_F=0.5t$, $N_F=70$; (c),(g) S/FM heterostructure with $\mu_F=0.5t$, $N_F=69$; (d),(h) S/FM heterostructure with $\mu_F=-t$, $N_F=70$. Other parameters are $h=0.3t$, $t_{SF}=0.3t$ for all FM cases. $N_S=100$, $\mu_S=t$ for all panels.} 
 \label{fig:h_eff_n_z}
\end{figure}

The dependence of $h_{\rm{eff}}$ on the branch number $n$ for the case of the insulating F layer is demonstrated in Fig.~\ref{fig:h_eff_n_z}(a). It is calculated according to the exact Eqs.~(\ref{kS_kF1}),(\ref{kS_kF2}) and (\ref{h_eff_def}). Nevertheless, the agreement with the analytical expression (\ref{h_eff_lim1}) is very good. The corresponding spatial distribution $h_{\rm{eff}}(z)$ over different superconducting sites, that is 
\begin{align}
    h_{\rm eff}(z)=\sum_n h_{\rm eff}(n) |\varphi_n(z)|^2,
\end{align}
is shown in Fig.~\ref{fig:h_eff_n_z}(b). In this expression $\varphi_n(z)=\sqrt{\frac{2}{N_S+1}}\sin\left(\frac{\pi n (a-z)/a}{N_S+1}\right)$ is an unperturbed wave function corresponding to branch $n$.  We can see that for the insulating F layer $h_{\rm{eff}}(z)$ is practically constant except for the boundary layers. Thus, we can conclude that for S/FI heterostructures the commonly accepted model of the S/F heterostructure as a homogeneous superconductor in some effective exchange field $h_{\rm{eff}}$ works well. 

The second limiting case, which can be considered analytically, is a tunnel interface, i.e. when $t_{SF}\ll t$. In this case for a given $n$ from Eqs.~(\ref{kS_kF1})-(\ref{kS_kF2}) one can obtain:
\begin{align}
    \Delta k^\sigma=\frac{2\sin(k_S^0d_S)(-1)^nt_{SF}^2 \sin(k_F^{0\sigma}d_F)}{(d_S+a)t^2\sin(k_F^{0\sigma}(d_F+a))},
    \label{eq:kS_lim2}
\end{align}
where $k_F^{0\sigma}(n)=\arccos(\cos(k_S^0(n)a)+(\mu_S-\tilde \mu_F^\sigma)/2t)$. Then
\begin{align}
    h_{\rm eff}(n)=\frac{-2 a \sin^2(\frac{\pi n}{N_S+1})t_{SF}^2 }{(d_S+a)t^2} \times~~~~~~~~ \nonumber \\ 
    \left(\frac{\sin(k_F^{0\uparrow}d_F)}{\sin(k_F^{0\uparrow}(d_F+a))}-\frac{\sin(k_F^{0\downarrow}d_F)}{\sin(k_F^{0\downarrow}(d_F+a))}\right) .
    \label{eq:h_eff_lim2}
\end{align}
This limiting case can be applied both for S/FI and S/FM heterostructures, but the main interest of the obtained expressions is for S/FM heterostructures. 
Three different examples of $h_{\rm{eff}}(n)$ for S/FM heterostructures are shown in Figs.~\ref{fig:h_eff_n_z}(b)-(d). They are calculated according to the exact Eqs.~(\ref{kS_kF1}),(\ref{kS_kF2}) and (\ref{h_eff_def}).  However, the main physical features of  $h_{\rm{eff}}(n)$ for S/FM heterostructures can be understood from analytical expression (\ref{eq:h_eff_lim2}). One can observe a very irregular behavior of $h_{\rm{eff}}(n)$. 
It occurs in energy regions where the conduction bands of the superconductor and ferromagnet intersect. In this case, as can be seen from Eq.~(\ref{eq:h_eff_lim2}), when moving to the next branch $n \to n+1$, $h_{\rm{eff}}$ can change sharply due to the oscillating dependence on $k_F^{0\sigma}(n)$. For the S/FM structure corresponding to Fig.~\ref{fig:h_eff_n_z}(d) the energy window, where the ferromagnet and superconductor conduction bands intersect, is narrower than for S/FM heterostructures corresponding to Figs.~(\ref{fig:h_eff_n_z})(b)-(c). For this reason the region of the irregular behavior of $h_{\rm{eff}}(n)$ in Fig.~\ref{fig:h_eff_n_z}(d) is also narrower as compared to Figs.~(\ref{fig:h_eff_n_z})(b)-(c). For the spectrum branches outside this region $n \gtrsim 55$ the ferromagnet exhibit insulating properties and the behavior of $h_{\rm{eff}}(n)$ is actually similar to Fig.~\ref{fig:h_eff_n_z}(a).

The corresponding spatial behavior of $h_{\rm{eff}}$ for the S/FM heterostructures corresponding to Figs.~(\ref{fig:h_eff_n_z})(b)-(d) is shown in Figs.~(\ref{fig:h_eff_n_z})(e)-(h), respectively. The irregular spatial dependence of $h_{\rm{eff}}(z)$ can have zero or nonzero spatial average. It is very sensitive to the particular parameters of the heterostructure and is unpredictable. In particular, $h_{\rm{eff}}(n)$ and $h_{\rm{eff}}(z)$ for S/FM heterostructures are highly sensitive to the number of ferromagnetic atomic planes. It can be seen by comparing Figs.~(\ref{fig:h_eff_n_z})(b) and (c), which only differ by this number: $N_F = 70$ for Fig.~(\ref{fig:h_eff_n_z})(b) and $N_F = 69$ for Fig.~(\ref{fig:h_eff_n_z})(c). However, the particular distribution of the exchange field as a function of $n$ or $z$ is completely different. Moreover, the averaged value of $h_{\rm eff}(z)$ [see panels Fig.~(\ref{fig:h_eff_n_z})(f),(g)] has the opposite sign.

\begin{figure}[tbh!]
\centering
\includegraphics[width=0.9\columnwidth]{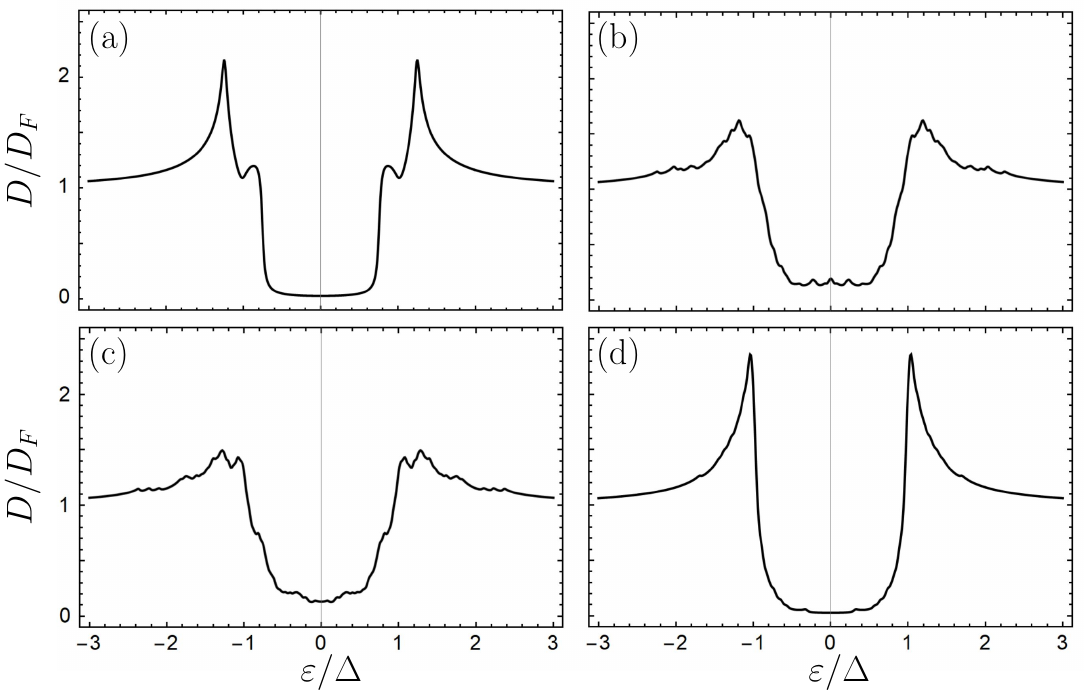}
\caption{LDOS calculated according to Eq.~(\ref{eq:LDOS_model}).  Panels (a)-(d) correspond to the S/F systems, for which $h_{\rm eff}(n)$ is represented in Figs.~\ref{fig:h_eff_n_z}(a)-(d), respectively. The Dynes parameter $\Gamma = 0.025\Delta$. $\Delta=0.002t$.} 
 \label{fig:LDOS_model}
\end{figure}

The LDOS calculated as 
\begin{align}
D(\varepsilon) = \sum_{n,\sigma} {\rm Im}[\frac{|\varphi_n(a-d_S)|^2 (\varepsilon+i\Gamma+h_{\rm eff}(n)\sigma)}{\sqrt{\Delta^2-(\varepsilon+i\Gamma+h_{\rm eff}(n)\sigma)^2}}] 
\label{eq:LDOS_model}
\end{align}
is presented in Fig.~\ref{fig:LDOS_model}. The Dynes parameter $\Gamma$ models the broadening of the LDOS peaks due to inelastic electronic scattering. Panels (a)-(d) of Fig.~\ref{fig:LDOS_model} correspond to the S/F systems, for which $h_{\rm eff}$ is represented in Figs.~\ref{fig:h_eff_n_z}(a)-(d), respectively. It is seen that for the S/FI system [see Figs.~\ref{fig:h_eff_n_z}(a),(e) and Fig.~\ref{fig:LDOS_model}(a)] the LDOS manifests well-pronounced spin splitting of the coherence peaks, which is well described by an effective model of the homogeneous superconductor in the Zeeman field $h_{\rm eff}(z)$, shown in Fig.~\ref{fig:h_eff_n_z}(e). 

LDOS for S/FM heterostructures does not manifest a spin splitting of the coherence peaks even if the effective exchange field has nonzero averaged value. It can have different shape depending on the average amplitude of $h_{\rm eff}$ oscillations. If the oscillation amplitude is rather small, as in Fig.~\ref{fig:h_eff_n_z}(d), the corresponding LDOS looks similar to the conventional BCS-type behavior {\it without a Zeeman splitting} of the coherence peaks [see Fig.~\ref{fig:LDOS_model}](d). However, if the amplitude of $h_{\rm eff}$ oscillations is high enough and exceeds the value of the superconducting gap $\Delta$, the LDOS can manifest gapless behavior with smeared coherence peaks [see Figs.~\ref{fig:LDOS_model}(b),(d)]. It is important that this behavior is not related to the suppression of superconductivity by the proximity effect with the ferromagnet. The value of the superconducting order parameter $\Delta$ is taken the same for all panels of Fig.~\ref{fig:LDOS_model}. The reason for the gap closing can be traced from Eq.~(\ref{eq:LDOS_model}). The spin-up (spin-down) BCS-like DOS for a given spectrum branch $n$ is shifted to the right (left) along the energy axis by $h_{\rm eff}(n)$. If $h_{\rm eff}(n)>\Delta$ the gap in the partial DOS for this branch is closed. The spin splitting of the net DOS is smeared due to chaotic distribution of $h_{\rm eff}(n)$ at different branches.

Thus, the analysis of this section indicates that the proximity effect with a ferromagnetic insulator creates a well-defined spin splitting of the electronic spectra in the superconductor, which is approximately uniform in space for thin superconducting films and leads to a well-defined Zeeman splitting of the LDOS. Therefore, S/FI heterostructures are well described by the model of a homogeneous superconductor in an effective exchange field. The proximity effect with a ferromagnetic metal also creates a spin splitting of the spectra in a superconductor. However, it has a chaotic spatial structure and does not lead to a clear spin splitting of the LDOS. Thus, the proximity effect in S/FM heterostructures cannot be described within the framework of the model of a homogeneous superconductor  in some effective exchange field.

The calculation performed is valid for ballistic systems in the absence of impurities and interface defects, such as roughness or the presence of terraces. The presence of impurities in the system will lead to the fact that different branches of the electron spectrum will interact and mix as a result of electron scattering on impurities. Thus, the electron will feel the exchange field averaged over several branches. As a result, the deviation from the averaged value of $h_{\rm{eff}}$ will decrease and in the limit of strong scattering, the electron will feel the average $h_{\rm{eff}}$.  Thus, qualitatively, the presence of diffusive scattering in the system should lead to the appearance of a more pronounced Zeeman splitting of the LDOS peaks, and in the diffusive limit the model of the homogeneous superconductor in the effective exchange field becomes applicable to S/FM heterostructures \cite{Bergeret2001,Bobkova2015}. In the regime of intermediate impurity concentration {\it superconductivity of the S/FM heterostructures can be enhanced upon increasing the impurity concentration} because in general the average $h_{\rm eff}$ should be rather small.  

The roughness of the S/FM interface affects the LDOS similarly, since diffusive reflection from the interface also leads to mixing of different branches of the spectrum. However, the presence of terraces on the interface can lead to the opposite effect. The fact is that, as shown in Figs.~\ref{fig:h_eff_n_z}(b) and (c), the chaotic effective field induced by a ferromagnetic metal is very sensitive to the number of atomic layers of the superconductor. Therefore, the presence of several terraces will lead to the superposition and averaging of several random effective field patterns, as a result of which the spin splitting of the LDOS will be smeared.

\section{Proximity-induced Zeeman splitting, triplet correlations and DOS in S/FI and S/FM thin-film heterostructures: numerical results}
\label{sec:numerical}

Here we investigate the influence of the proximity-induced spin splitting on the superconducting properties and LDOS in S/F heterostructures.  In this case Eqs.~(\ref{kS_kF1}) and (\ref{kS_kF2}) are no longer valid. Therefore, we numerically solve the Bogolubov-de Gennes (BdG) equations, which can be obtained from  Hamiltonian (\ref{eq_hamBDG}) by exploiting the Bogolubov transformation:
\begin{align}
\hat \psi_{\bm i\sigma}=\sum\limits_n u^{\bm i}_{n\sigma}\hat b_n+v^{\bm i*}_{n\sigma}\hat b_n^\dagger . 
\label{bogolubov}
\end{align}
The resulting BdG equations take the form:
\begin{align}
&-\mu u^{\bm i}_{n,\sigma}-t\sum \limits_{\bm j \in \langle \bm i \rangle}u_{n,\sigma}^{\bm j} + \sigma \Delta_{\bm i} v^{\bm i}_{n,-\sigma}+ \nonumber \\
&+\sum_\alpha(\bm {h}_{\bm i} \bm{\sigma})_{\sigma\alpha}u_{n,\alpha}^{\bm i}  = \varepsilon_n u_{n,\sigma}^{\bm i} \nonumber \\  
&-\mu v^{\bm i}_{n,\sigma}-t \sum \limits_{\bm j \in \langle \bm i \rangle}v_{n,\sigma}^{\bm j}+ \sigma \Delta_{\bm i}^* u^{\bm i}_{n,-\sigma}+\nonumber\\ 
&+\sum_\alpha(\bm{h}_{\bm i}\bm{\sigma}*)_{\sigma\alpha}v_{n,\alpha}^{\bm i}  = -\varepsilon_n v_{n,\sigma}^{\bm i}, \label{bdg}
\end{align}
where $\langle \bm i \rangle$ means nearest neighbors of the site $\bm i$. The BdG equations are to be solved together with the self-consistency equation:
\begin{align}
\Delta_{\boldsymbol{i}} &= \lambda\langle\hat \psi_{\bm{i} \downarrow} \hat \psi_{\bm{i} \uparrow} \rangle \nonumber \\
&=  \lambda \sum\limits_{n} (u_{n,\downarrow}^{\bm i} v_{n,\uparrow}^{\bm i*}(1-f_n)+u_{n,\uparrow}^{\bm i} v_{n,\downarrow}^{\bm i*}f_n).
\label{eq:self-consist_BDG}
\end{align}

\begin{figure}[tbh!]
\centering
\includegraphics[width=0.9\columnwidth]{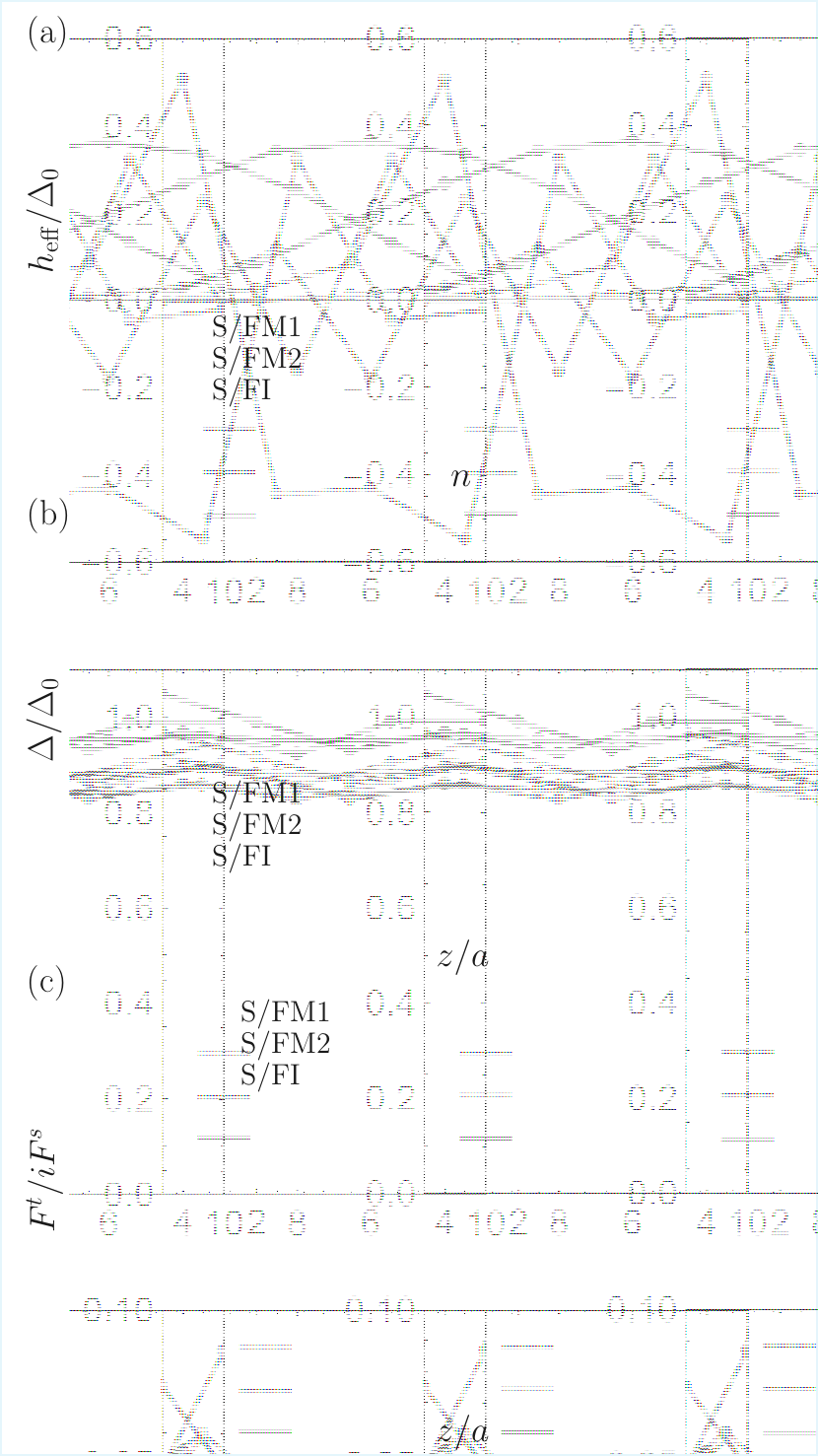}
\caption{Spatial distribution of the main proximity-induced properties across the S layer. Three different S/F bilayers are represented: S/FI bilayer and two S/FM bilayers denoted as S/FM1 and S/FM2. (a) Effective exchange field $h_{\rm eff}$; (b) superconducting order parameter $\Delta$; (c) triplet correlations at first Matsubara frequency $F^t(\omega_1)$. $h_{\rm eff}$ and $\Delta$ are measured in units of the superconducting OP of the isolated S layer $\Delta_0$ taken at the same temperature. $F^t$ is measured in units of the singlet superconducting correlations $F^s(\omega_1)$. The parameters of the considered systems are the following. S/FI: $\mu_F=-10t$, $t_{SF}=t$, $h=4t$. S/FM1: $\mu_F=-t$, $t_{SF}=0.2t$, $h=0.5t$. S/FM2: $h=1.5t$, the other parameters are the same as for S/FM1. For all considered systems $\mu_S=t$, $\Delta_0=0.0195t$, $T=0.051\Delta_0$.} 
 \label{fig:triplets_bilayer}
\end{figure}

Fig.~\ref{fig:triplets_bilayer}(a) shows $h_{\rm{eff}}(n)$  calculated according to Eq.~(\ref{h_eff_def}) for three different S/F bilayers in the superconducting state. Since we perform a direct numerical diagonalization of the BdG equations, we can only consider rather small numbers of the superconducting and ferromagnetic atomic layers. In particular, in Fig.~\ref{fig:triplets_bilayer} the results corresponding to $N_S = N_F = 10$ are demonstrated. Nevertheless, the presented numerical results manifest the same behavior as the results obtained from Eqs.~(\ref{kS_kF1}) and (\ref{kS_kF2}) and shown in Figs.~\ref{fig:h_eff_n_z}(a)-(d). For the S/FI heterostructure [green curve in Fig.~\ref{fig:triplets_bilayer}(a)] the dependence $h_{\rm{eff}}(n)$ is smooth resulting in nearly spatially constant $h_{\rm{eff}}$ across the S layer. Red and blue curves in Fig.~\ref{fig:triplets_bilayer}(a) correspond to two different S/FM bilayers: S/FM1 bilayer has smaller true exchange field of the FM layer $h_1 = 0.5 t$, and for S/FM2 bilayer the true exchange field of the FM layer is higher, $h_2 = 1.5 t$. For both S/FM systems one can see irregular oscillating behavior of $h_{\rm{eff}}$. The other important fact is that in spite of strongly different values of the true exchange field $h_2/h_1 = 3$ the amplitude of effective exchange field is very similar for both S/FM bilayers. It is in agreement with Eq.~(\ref{eq:h_eff_lim2}), which indicates that for S/FM heterostructures there is no proportionality between $h$ and $h_{\rm eff}$, in contrast to the case of S/FI heterostructures described by Eq.~(\ref{h_eff_lim1}).

To investigate the triplet correlations that are induced in the S layer due proximity to the F layer, one needs to calculate the anomalous Green's function, which actually contains information about both singlet and triplet superconducting correlations that arise at the interface. The anomalous Green's function in Matsubara representation can be calculated as $F_{\bm i, \alpha \beta} = - \langle T_\tau \hat \psi_{\bm i \alpha}(\tau) \hat \psi_{\bm i \beta}(0) \rangle$, where $\tau$ is the imaginary time. The component of this anomalous Green's function for a given Matsubara frequency $\omega_m = \pi T(2m+1)$ in the framework of the BdG approach is calculated as follows: 
\begin{align}
F_{\bm i,\alpha\beta}(\omega_m)= \sum\limits_n (\frac{  u_{n,\alpha}^{\bm i} v_{n,\beta}^{\bm i*}}{i \omega_m -\varepsilon_n}+\frac{ u_{n,\beta}^{\bm i} v_{n,\alpha}^{\bm i*}}{i \omega_m +\varepsilon_n}).
\label{eq:anom_func_BDG}
\end{align}
For the considered case of a homogeneous ferromagnetic order only off-diagonal in spin space components, corresponding to opposite-spin pairs, are nonzero. The singlet (triplet) correlations are described by $F_{\bm i}^{s,t}(\omega_m) = F_{\bm i,\uparrow \downarrow}(\omega_m) \mp F_{\bm i,\downarrow \uparrow}(\omega_m)$. It is important to note that on-site triplet correlations are odd in Matsubara frequency, as it should be according to the general fermionic symmetry. 

Due to the translational invariance along the S/F interface the superconducting OP and anomalous Green's function do not depend on coordinate along the interface. The spatial distribution of the superconducting OP $\Delta (z)$ and the triplet correlations at first Matsubara frequency $F^t(\omega_1,z)$ along the normal $z$ to the interface  in the S layer is presented in Figs.~\ref{fig:triplets_bilayer}(b) and (c), respectively. The coordinate dependence of $\Delta$ is very weak in agreement with the fact that the thickness of the S layer $N_S=10$ is smaller than the superconducting coherence length $\xi_S = v_F/\Delta\approx 100 a$. The spatial variation of the triplet correlations is also rather weak. The overall amplitude of $F^t$ for S/FM1 and S/FM2 bilayers is also similar in spite of very different values of the true exchange field of the ferromagnets, as it is observed for $h_{\rm eff}$ in Fig.~\ref{fig:triplets_bilayer}(a). 

Next, by the example of S/FM1 and S/FM2 heterostructures, we demonstrate that, due to the irregular structure of $h_{\rm eff}$, the LDOS in S/FM heterostructures with close values of triplet correlations may or may not exhibit splitting of the coherent peaks, which is widely considered as a signature of the presence of the proximity-induced spin splitting of the electronic spectra and, most importantly, triplet correlations in the S layer. 

\begin{figure}[tbh!]
\centering
\includegraphics[width=0.9\columnwidth]{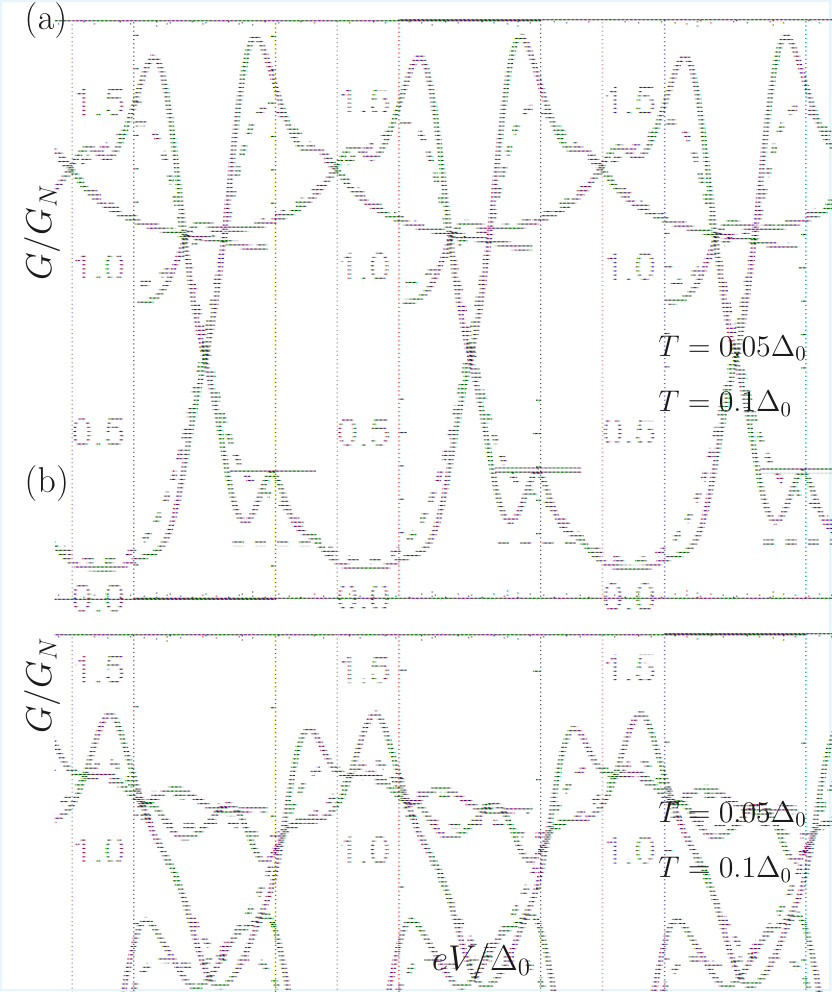}
\caption{Conductance $G$ of the tunneling junctions between the impenetrable superconducting surface of edge of the S/FM1 (a) and S/FM2 (b) bilayers and a normal probe. $G$ is normalized to the conductance of the same systems in the normal state. The parameters are the same as in Fig.~\ref{fig:triplets_bilayer}.} 
 \label{fig:LDOS_bilayer10}
\end{figure}

The LDOS at the impenetrable surface of the S/F bilayer can be investigated by measuring the tunneling conductance between the bilayer and a normal or a superconducting electrode. In Fig.~\ref{fig:LDOS_bilayer10} we represent the low-temperature conductance $G$ of the tunneling junction between the S layer of the S/FM1 and S/FM2 bilayers and a normal electrode, which is related to the LDOS at the impenetrable surface of the S layer as follows:
\begin{align}
G = \int d\varepsilon D_N(\varepsilon)D(\varepsilon)\frac{4}{T {\rm cosh}^2\frac{\varepsilon+eV}{2T}},
\label{conductance}
\end{align}
where $D_N(\varepsilon)$ is the LDOS of the normal probe electrode. The Zeeman splitting of the LDOS coherence peaks is clearly seen for the S/FM2 heterostructure. At the same time, the S/FM1 heterostructure does not demonstrate the clear Zeeman splitting even at very low temperatures. The standard BCS-type density of states is superimposed by small oscillations from which it is impossible to extract any Zeeman splitting. In both cases we observe a strong broadening of the coherent peaks and a non-zero density of states for subgap energies. Usually these characteristic features are described by a pair-breaking factor, which can have various physical origins, including the orbital effect of the magnetic field \cite{Maki1964},  inelastic scattering processes, spin-orbit scattering \cite{Fulde1966} or the effect of magnetic impurities \cite{Abrikosov1961}. However, it is worth emphasizing once again that in the case we are considering, the broadening and the subgap DOS are not associated with the above-mentioned reasons, since they are simply not included in the model, but is determined by the random distribution of $h_{\rm eff}(n)$ along the branches of the spectrum with a wide range of values. Thus, this is also worth keeping in mind when analyzing the possible reasons of the broadening of the LDOS in S/FM heterostructures.  With irregular behavior of $h_{\rm eff}(n)$, which is typical for S/FM heterostructures, the presence or absence of splitting of coherent peaks is determined by fine details of the specific $h_{\rm eff}(n)$ distribution and cannot be predicted. In general, with an increase in the number of superconductor layers, the probability of observing the clearly defined splitting tends to zero due to the increasing randomization of the $h_{\rm eff}(n)$ distribution. 

\begin{figure}[tbh!]
\centering
\includegraphics[width=0.9\columnwidth]{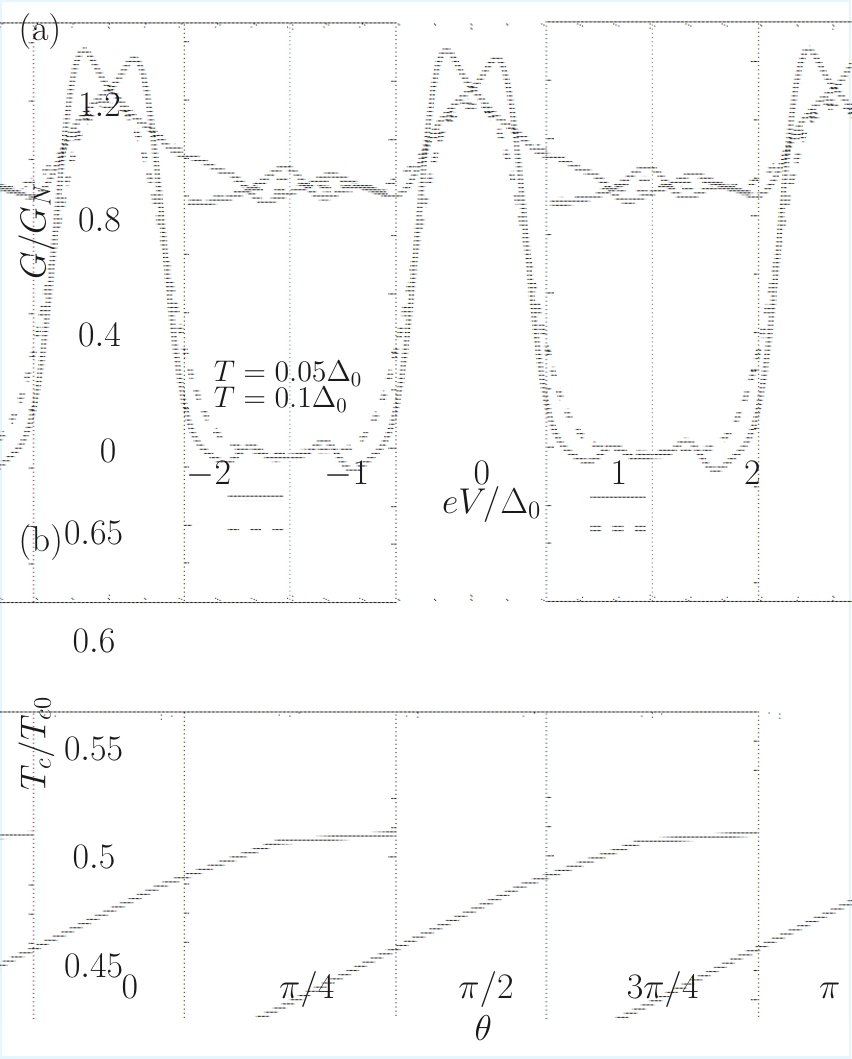}
\caption{(a) Conductance $G$ of the tunneling junction between the impenetrable superconducting surface of the S/FM bilayer and a normal probe. (b) Superconducting critical temperature of the FM/S/FM trilayer as a function of the angle $\theta$ between the magnetic moments of the FM layers. The trilayer differs from the bilayer only by the addition of a left FM layer identical to the right one. $N_S=20$, $N_F=10$, $\Delta_0=0.0195t$, $h=0.5t$, $t_{SF}=0.4t$, $\mu_F=-t$, $\mu_S=t$, $T=0.051\Delta_0$.} 
 \label{fig:valve}
\end{figure}

However, as it is demonstrated by examples of S/FM1 and S/FM2 heterostructures, the absence of the Zeeman splitting in the LDOS does not mean the absence of proximity-induced triplet correlations in S/FM heterostructures. Thus, such heterostructures can be successfully used in superconducting spintronics because it is precisely triplet correlations that are the fundamental basis of many effects of superconducting spintronics.  In particular, below we demonstrate that the superconducting spin valve effect can be observed in S/FM heretostructures even if the LDOS does not manifest the Zeeman splitting of the coherence peaks. 

The low-temperature tunneling conductance between a S/FM bilayer with $N_S=20$ superconducting atomic layers and a normal probe electrode is shown in Fig.~\ref{fig:valve}(a). The superconducting critical temperature of the corresponding FM/S/FM trilayer as a function of the angle $\theta$ between the magnetic moments of the FM layers is plotted in Fig.~\ref{fig:valve}(b). All the parameters of the S layer and the FM layers are the same for the bilayer and trilayer structures. Similar to the results represented above the low-temperature conductance manifest broadened and distorted coherence peaks with no clear Zeeman splitting, see Fig.~\ref{fig:valve}(a). At the same time, for the FM/S/FM trilayer $T_c(\theta=\pi)-T_c(\theta=0)/T_c(\theta=\pi) \approx 0.2$. That is, the trilayer system manifests 20\% spin-valve effect. The reason is that in spite the irregular spatial profile of $h_{\rm eff}$ and the triplet correlations in the S layer of the heterostructure, their averaged over the S layer thickness values depend on the mutual magnetization orientations of the FM layers. For $\theta = 0$ the averaged $h_{\rm eff}$ induced by both FM layers are summed up, thus resulting in more suppressed superconductivity. At the same time for $\theta = \pi$ they partially cancel each other and the suppression of superconductivity is weaker. 

Thus, the mechanism of the spin valve effect is actually the same as was discussed long ago by de Gennes \cite{DEGENNES1966} for ferromagnetic insulators, although for FM/S/FM structures with ferromagnetic metals, instead of the spatially homogeneous effective exchange field considered in \cite{DEGENNES1966}, one should think about the average effective exchange field $\bar h_{\rm eff}(z)$. In general, $\bar h_{\rm eff}(z)$ in FM/S/FM heterostructures should decrease as $\propto 1/\sqrt{N_S}$ with an increase in the number of superconductor atomic layers, but this should not be an obstacle to the implementation of the spin-valve effect in FM/S/FM heterostructures, since for effective suppression of superconductivity, and therefore the operation of the spin valve, small effective exchange fields of the order of $\Delta$ are sufficient. 

It is worth mentioning that in contrast to thin-film  S/FI heterostructures, where practically all spintronics effects can be described in terms of $h_{\rm eff}$,  for S/FM heterostructures  $\bar h_{\rm eff}(z)$ is not a universal characteristics  describing the proximity effect. This value is well suited for a qualitative description of the physics of the spin-valve effect, but is not relevant for other important effects of superconducting spintronics, in particular transport effects. For example, the result of calculating the thermally induced spin current and the giant spin-dependent Seebeck effect \cite{Bergeret2018,Heikkila2019,Bobkova2021,Machon2013,Ozaeta2014,Machon2014,Kolenda2016,Kolenda2017} in S/FM heterostrucutres will be determined not by $\bar h_{\rm eff}(z)$, but by some effective value of $h_{\rm eff}$ averaged over all branches of the spectrum with weighting coefficients proportional to the projection of the Fermi velocity for a given branch of the spectrum onto the direction of the current. However, the detailed study of this problem is beyond the scope of the present work.

\section{Proximity effect in S/AM heterostructures}
\label{sec:numerical_AM}

In principle, the difference between the proximity effect in superconducting heterostructures with metals and insulators is not limited to the case of ferromagnets and is also relevant for other types of magnets. As an example, here we investigate the proximity effect in thin-film superconductor/altermagnet heterostructures with altermagnetic metals (S/AM) and insulators (S/AI). The appropriate Bogolubov-de Gennes equations take the form of Eq.~(\ref{bdg}) with the substitution $\sum \limits_\alpha (\bm h_{\bm i} \bm \sigma)_{\sigma \alpha} u_{n,\alpha}^{\bm i} \to \sum \limits_{\bm j \in \langle \bm i \rangle, \alpha} (\bm h_{\bm i \bm j} \bm \sigma)_{\sigma \alpha} u_{n,\alpha}^{\bm j}$ and with the analogous substitution for the term containing $\bm \sigma^*$.

\begin{figure}[tbh!]
\centering
\includegraphics[width=0.8\columnwidth]{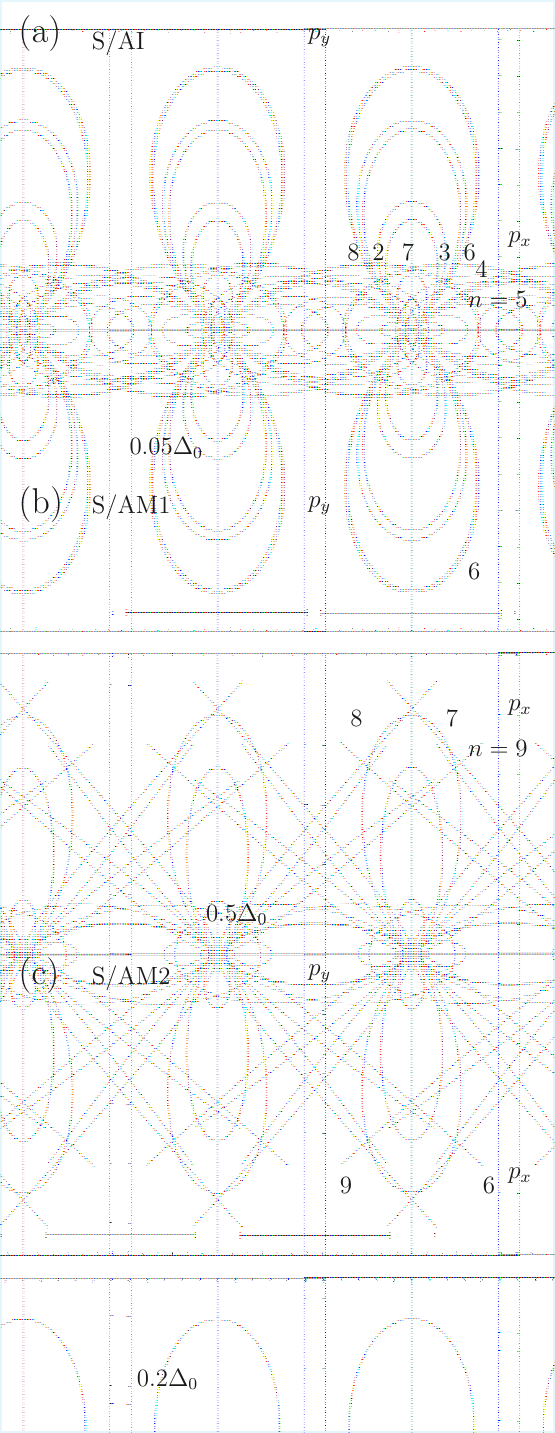}
\caption{Spin splitting $h_{\rm{eff}}(n,\bm p,\varepsilon)$ of the electronic spectra of (a) the S/AI bilayer  and (b)-(c) two different S/AM bilayers at $\varepsilon = 3t$.  The sign of the spin splitting is indicated by color:  $h_{\rm eff}>0$ is in red and $h_{eff}<0$ is in blue. The absolute value of the spin splitting for a given momentum direction in the BZ is shown as a distance from the origin. The corresponding scale is indicated on each panel.  The branch number $n$ is indicated by a number. The parameters of the considered systems are the following. S/AI: $\mu_F=-10t$, $t_{SF}=t$, $h=0.1t$. S/AM1: $\mu_F=-t$, $t_{SF}=0.1t$, $h=0.1t$. S/AM2: $\mu_F=-t$, $t_{SF}=0.1t$, $h=0.01t$. For all considered systems $\mu_S=t$, $\Delta_0=0.00914t$, $T=0.11\Delta_0$.} 
 \label{fig:splitting_AM}
\end{figure}

Fig.~\ref{fig:splitting_AM} shows spin splitting of the electronic spectra of the S/AI bilayer [Fig.~\ref{fig:splitting_AM}(a)] and the S/AM bilayer [Fig.~\ref{fig:splitting_AM}(b)-(c)]. In this case the spin splitting on each of the discrete electronic branches  depends on the momentum direction in the BZ and on the energy, that is $h_{\rm{eff}} = h_{\rm{eff}}(n,\bm p,\varepsilon)$. The sign of the spin splitting is indicated by color:  $\varepsilon_{n,\uparrow} - \varepsilon_{n,\downarrow}>0$ is in red and $\varepsilon_{n,\uparrow} - \varepsilon_{n,\downarrow}<0$ is in blue. The absolute value of the spin splitting for a given momentum direction in the BZ is shown as a distance from the origin. 

It is seen that the antisymmetry with respect to the rotation over $\pi/2$ in the BZ, which is dictated by the exchange field of the altermagnet, is preserved for all considered heterostructures. However, pure $d$-wave symmetry inherent in the true exchange field of the considered altermagnet is observed only for S/AI heterostructures, and for S/AM heterostructures higher harmonics also appear. Moreover, if we focus on the absolute value of the spin splitting, for example $h_{\rm eff}(n,p_y=0,\varepsilon)$, we can conclude that it exhibits the same general trends as $h_{\rm eff}(n)$ for S/F heterostructures. Namely, the dependence  is smooth for S/AI heterostructures resulting in nearly spatially constant $h_{\rm{eff}}$ across the S layer. At the same time for both S/AM systems one can see irregular oscillating behavior of $h_{\rm{eff}}(n, p_y=0, \varepsilon)$. 

\begin{figure}[tbh!]
\centering
\includegraphics[width=0.8\columnwidth]{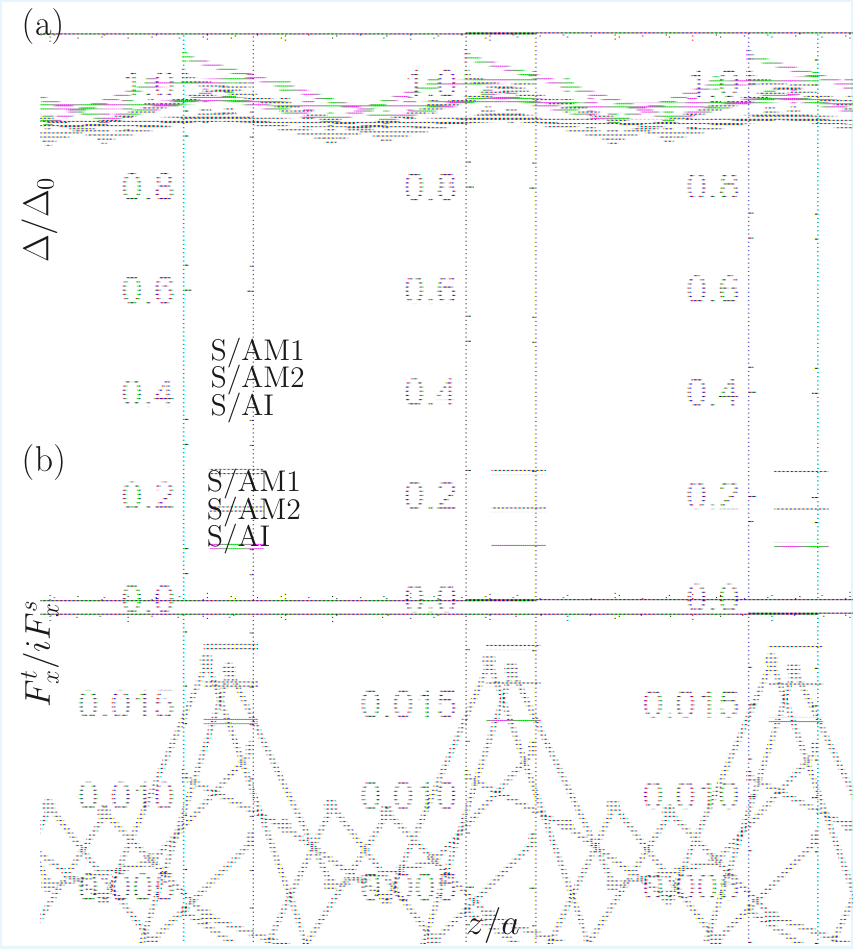}
\caption{(a) Superconducting order parameter $\Delta$; (b) triplet correlations at first Matsubara frequency $F_x^t(\omega_1)$. The parameters of the considered systems are same as in Fig.~\ref{fig:splitting_AM}} 
 \label{fig:triplets_AM}
\end{figure}

The spatial distribution of the superconducting OP $\Delta (z)$ and the triplet correlations at first Matsubara frequency $F_x^t(\omega_1,z) \equiv F^t(\omega_1,z, p_y=0)$ along the normal $z$ to the interface  in the S layer are presented in Fig.~\ref{fig:triplets_AM} for all three considered systems. In full analogy with the case of S/F heterostructures, the coordinate dependence of $\Delta$ is very weak. It is in agreement with the fact that the thickness of the S layer $N_S=10$ is smaller than the superconducting coherence length $\xi_S = v_F/\Delta\approx 200 a$. For the S/AI bilayer the spatial variation of the triplet correlations is also rather weak. Also for this case the spin splitting of the electronic spectra in the S layer preserves the $d$-wave symmetry, smooth distribution over the discrete electronic branches and nearly constant spatial behavior across the S layer. It means that the thin-film S/AI bilayers can be treated in the framework of the model of the homogeneous superconductor in the presence of some effective exchange field having $d$-wave momentum symmetry.   

At the same time, the amplitude of triplet correlations $F_x^t$ for S/AM1 and S/AM2 bilayers irregularly oscillates across the S layer due to the irregular structure of $h_{\rm eff}(n)$. It is the same behavior as was demonstrated above for S/FM thin-film bilayers.Therefore, this result once again demonstrates the inapplicability of the model of a homogeneous superconductor in an effective exchange field to describe the proximity effect in heterostructures with metallic magnets of various types.

\section{Conclusions}
\label{sec:conclusions}

A comparative analysis of the proximity effect in thin-film ballistic S/FM and S/FI heterostructures is performed. It is shown that in the S/FI heterostructures the proximity effect
creates a well-defined spin 
splitting of the electronic spectra in the S layer, which is a smooth function of the spectral branch and leads to a well-defined
Zeeman splitting of the LDOS. Thus, S/FI heterostructures are well described by the model of a homogeneous
superconductor in an effective exchange field, which is directly proportional to the true exchange field of the ferromagnet. The proximity effect in S/FM heterostructures also creates a spin
splitting of the spectra of the S layer.  However, this spin splitting is an irregular function of the spectrum branch, what also results in chaotic spatial distribution of the effective exchange field across the S layer. For this reason S/FM heterostructures in general does not demonstrate a clear spin splitting of the LDOS. Moreover, the amplitude of the chaotic effective exchange field is not proportional to the true exchange field. It is very sensitive to fine details of the materials and geometry of the heterostructure and in fact is unpredictable. Thus, the proximity effect in S/FM heterostructures cannot be described
within the framework of the model of a homogeneous superconductor in some effective exchange field. Nevertheless, in spite of the irregular and unpredictable character of the spin splitting in the S layer and impossibility to detect it via the spin splitting of the LDOS, the S/FM heterostructures support well-pronounced triplet correlations and, thus, can be used for spintronics applications. As an example, we demonstrate the 20\% spin-valve effect in FM/S/FM trilayer structures with no clear spin splitting in the LDOS. 

An analogous comparative analysis of the proximity effect in superconductor/altermagnet heterostructures with insulating and metallic $d$-wave altermagnets is also performed. The first important conclusion is that the spin-splitting induced in the S layer due to proximity to the altermagnet preserves the antisymmetry with respect to $\pi/2$-rotation in the BZ regardless of whether the altermagnetic material is an insulator or a metal. However, for S/AM heterostructures higher harmonics in the momentum dependence of the spin splitting generally appear. The second conclusion is that $h_{\rm eff}(n)$ for the S/AM heterostructures manifests all the key features reported for S/F heterostructures. Namely, S/AM heterostructures with insulating altermagnets are well described by the model of a homogeneous
superconductor in a $d$-wave effective exchange field, which is directly proportional to the true exchange field of the ferromagnet. For S/AM heterostructures with metallic altermagnets the $d$-wave spin splitting is again an irregular function of the spectrum branch, what results in chaotic spatial distribution of $h_{\rm eff}$ across the S layer. Thus, in full analogy with S/FM heterostructures the model of a homogeneous superconductor in some effective exchange field is not applicable to the proximity effect in S/AM heterostructures with metallic altermagnets.

The model of a homogeneous superconductor in an effective exchange field is often used to study various effects in superconductor/magnet heterostructures. In particular, this model was recently applied to superconductor/altermagnet heterostructures to study the effect of nonmagnetic impurities on superconductivity \cite{Vasiakin2025}. It was predicted that the critical temperature of such heterostructures increases with increasing nonmagnetic impurity concentration. This occurs because scattering by impurities averages the momentum-dependent effective exchange field experienced by the electron. The path-averaged effective exchange field becomes smaller, and the critical temperature, suppressed by this effective field, increases. Our results show that these conclusions about the influence of impurities on the superconducting state are fully applicable to S/AI heterostructures due to the applicability of the homogeneous superconductor model to describe them.

At the same time, in S/AM heterostructures the effect of superconductivity recovering by impurities should be enhanced compared to the results of the effective homogeneous model. This is explained by the fact that impurity scattering allows electrons to hop between different branches of the electron spectrum. As a result, the effective exchange field $h_{\rm eff}(n)$ experienced by the electron is averaged over different $n$. Since in heterostructures with metallic magnets this field depends chaotically on $n$, averaging leads to its significant weakening. As a result, superconductivity, suppressed by this field, is restored as the impurity concentration increases. From the above qualitative considerations, it also follows that even thin-film heterostructures with conventional ferromagnetic metals, and not just altermagnets \cite{Vasiakin2025} and antiferromagnets \cite{Bobkov2022,Bobkov2023}, should exhibit superconductivity restoration as the impurity concentration increases.

\begin{acknowledgments}
G.A.B. and I.V.B. acknowledge the support from Theoretical Physics and Mathematics Advancement Foundation “BASIS” via the project No. 23-1-1-51-1. The numerical calculations of superconducting DOS were supported by the Russian Science Foundation via the RSF project No. 24-12-00152. The analytical calculations of $h_{\rm eff}$ have been obtained under the support of MIPT via the project FSMG-2023-0014.
\end{acknowledgments}

\bibliography{DOS_SF}

\end{document}